\documentclass[sigconf]{acmart}
\usepackage{booktabs} 
\usepackage{tabularx}
\usepackage[ruled]{algorithm2e}
\usepackage{amsmath}
\usepackage{amssymb}
\usepackage[per-mode=symbol-or-fraction]{siunitx}
\newcommand{\ETX}{\textsf{ETX}}
\newtheorem{Definition}{Definition}

\newtheorem{Remark}{Remark}
\setcopyright{none}
\renewcommand\footnotetextcopyrightpermission[1]{}
\begin{document}
\title{Antilizer: Run Time Self-Healing Security for Wireless Sensor Networks}

\author{Ivana Tomi\'c, Po-Yu Chen, Michael J. Breza and Julie A. McCann}
\affiliation{%
  \institution{Imperial College London, Department of Computing}
  \streetaddress{P.O. Box 1212}
  \city{London}
  \country{United Kingdom}
}
\email{{i.tomic, po-yu.chen11, michael.breza04, j.mccann}@imperial.ac.uk}








\renewcommand{\shortauthors}{I. Tomi\'c et al.}

\begin{abstract}
Wireless Sensor Network (WSN) applications range from domestic Internet of Things systems
like temperature monitoring of homes to the monitoring and control of large-scale critical
infrastructures.
The greatest risk with the use of WSNs in critical infrastructure is their vulnerability to 
malicious network level attacks.
Their radio communication network can be disrupted, causing them to lose or delay
data which will compromise system functionality.
This paper presents Antilizer, a lightweight, fully-distributed solution to enable WSNs to
detect and recover from common network level attack scenarios.
In Antilizer each sensor node builds a self-referenced trust model of its neighbourhood
using network overhearing.
The node uses the trust model to autonomously adapt its communication decisions.
In the case of a network attack, a node can make neighbour collaboration routing decisions
to avoid affected regions of the network.
Mobile agents further bound the damage caused by attacks.
These agents enable a simple notification scheme which propagates collaborative decisions
from the nodes to the base station.
A filtering mechanism at the base station further validates the authenticity of the
information shared by mobile agents.
We evaluate Antilizer in simulation against several routing attacks.
Our results show that Antilizer reduces data loss down to 1\% (4\% on average), with
operational overheads of less than 1\% and provides fast network-wide convergence.
\end{abstract}

%
%




\maketitle

\section{Introduction} \label{sec:intro}


Wireless Sensor Networks (WSNs) are systems that consist of many small,
resource constrained sensor nodes.
WSNs have been successfully applied to many Internet of Things applications
providing ubiquitous connectivity and information-gathering capabilities
\cite{Milenkovic2006, Pascale2012}.
The natural evolution of WSNs is to make them part of larger systems.
These systems use WSN data as input to other processes.
Example systems include smart infrastructure systems such as water/waste
distribution networks, precision agriculture farms, and energy distribution
grids \cite{Rajkumar2010,Mo2012}.


The risk of the inclusion of WSNs into larger systems is that their operational
environment is often open to the public, and difficult to secure.
WSNs rely on radio networks that are easy to disrupt and subvert.
This makes them a potential target for cyber-attacks \cite{Kavitha2010}.
Attacks can be simple radio level attacks such as jamming, or more sophisticated
network level attacks where one or more sensor nodes are compromised and made to
behave in a malicious manner.
Attacks can cause data loss and increase data collection latency which will disrupt
the functioning of the system that relies upon data collected by the WSN \cite{Tomic2017}.


Extensive research efforts have been put into hardening WSN network protocols
with the use of various cryptographic mechanisms and pairwise key sharing
schemes (e.g. \cite{Perrig2002, Zhu2003, Karlof2004}) or building intrusion
detection schemes (e.g. \cite{Banerjee2005, daSilva2005, Krontiris2008}).
These approaches do not provide a mechanism for a WSN to recover from malicious
intrusion nor prevent disruption to the WSN or the application relying on data
collected by the WSN.  Recent attacks such as Mirai \cite{Kolias2017} have
shown the very real danger of WSNs being attacked by rouge nodes and that
this problem has not yet been effectively solved.
The challenge lies in the fact that the severe resource constraints and
uncontrolled operational environments of sensor nodes weaken the effectiveness
of current state-of-the-art security techniques.

Very few security approaches address both intrusion detection and autonomous
intrusion prevention for WSNs \cite{Sultana2014,Lu2015, Samian2017}.
The few that do, rely on an evaluation of a node's behaviour from either locally
or globally collected information. 
The global information approach is done at the base-station which creates
scalability limits and provides a single point of failure to an attacker.
The local information approach is done at the node level and combines surveillance 
techniques, such as overhearing and probing, and collaborations among the nodes
such as voting.
The greatest weakness of the local approach is that the information shared by
collaborating nodes can be easily falsified.
We argue that a node 'trusting data from its neighbours' introduces an additional
attack vector to be exploited.
A much more robust security approach should only use information that is collected,
and therefore trusted, by the node itself.
The challenges are how to interpret potentially noisy data and successfully categorise
malicious events from routine network changes.


In this paper, we present Antilizer, a run-time security solution for WSNs
that is able to detect network level attacks and at the same time adapt its
communication decisions to avoid the affects of the detected attack. 
Antilizer utilizes a self-referenced trust model at each node to evaluate
the behaviour of its one-hop neighbours.
Neighbour communication information is self-collected via network overhearing
which is a data collection method that is difficult to falsify.
This allows us to collect network metrics by counting the number of 
transmissions, receptions and other communication events without using the 
content of the communication.
The network metrics are mapped to a trust value using a kernel-based technique.
This notion of trust is used by a node when it makes communication decisions
regarding node collaboration (e.g. data routing, data aggregation) or
environmental awareness (e.g. adaptive duty cycling based on received signal
strengths).
In the example of data routing, self-referenced trust ensures that communication
avoids affected areas to prevent its loss during a network level attack.  
The main difference between our self-referenced trust model and existing 
trust-based models is that ours depends only on self-collected information, rather
than on potentially dishonest information provided by other nodes.

Upon detection of a malicious neighbour, there is the need to send the position
of the malicious node to the base station.
We use an agent-based notification scheme that introduces ANTs (Antilizer Notification
Tickets).
The ANT travels to
the base station and informs nodes along the way that changes in the network
behaviour are the result of malicious activity.
The ANTs reduce the number of false positive detections and constrain the damage of the
attack to a single neighbourhood in the network.
The authenticity of the information carried by the ANT is verified at the base 
station via a filtering mechanism.

The contributions of the paper are as follows:
\begin{enumerate}
\item A \textbf{self-referenced trust model} which enables each node to build knowledge
about its neighbourhood using only self-collected information and map this
knowledge to a trust model of its neighbourhood using a kernel-based approach
to generate a trust value for each neighbour.
\item An \textbf{agent-based notification scheme} which distributes information 
in the network to ensure the normal network operation upon the detection of an attack.
Our scheme overcomes the problem of distinguishing between genuine and malicious
network level changes and provides a global view of network behaviour essential
for complete network recovery. 
\item A \textbf{filtering mechanism at the base station} which verifies the
authenticity of the information distributed by the agent-based notification scheme.
\item An \textbf{implementation} of Antilizer in the Contiki operating system
\cite{Contiki}.
Its effectiveness and efficiency is evaluated in the case of node
collaboration for secure data routing.
\end{enumerate}

Antilizer is agnostic to WSN operating system and routing layers and achieves
low overheads of less than $1$\% on average and a detection reliability of
$99.3$\%.
It was evaluated on various sized networks, against
different attack scenarios and at a range of attack intensities.
Antilizer does not require off-line processing or training, provides a guarantee of zero
performance penalty in the presence of no attack, and is an inspiration for
further exploration of the use of learning-based methods on sensor nodes.

The remainder of this paper is organized as follows: Sec.~2 surveys the related
work.
Sec.~3 discusses the system model and our assumptions.
Sec.~4 gives an overview of Antilizer.
The design details are given in Sec.~5, Sec~6 and Sec.~7.
In Sec.~8 and Sec.~9 we present the implementation and evaluation of Antilizer,
respectively.
We end the paper in Sec.~10 with brief concluding remarks.

\section{Related work}

Antilizer is a combination of two broad categories of security systems for WSNs,
trust-based and automated response systems.

\textbf{Trust-based security schemes.}
There is a large body of theoretical and practical results for WSN trust-based
security schemes.
These use the approach of monitoring neighbours for behavioural anomalies to
achieve reliable network communication \cite{Yan2014, Guo2017}.
We only discuss schemes that use network metrics in a similar way to Antilizer.

Trust-based methods such as \cite{Tajeddine2012} work in a centralized manner
which requires a global view of the network.
All of the network information for each node has to be passed to the base
station for processing.
This approach assumes that the data can be safely sent to and returned from
the base station and that the base station can handle the processing for the
entire network.
Antilizer does not have these limitations because each node creates its own
trust model of its neighbours.
This fully distributed approach improves scalability, lowers energy
consumption and makes the solution less vulnerable to malicious activity.

Fully distributed schemes \cite{Chen2011,Karkazis2012,Bao2012,Chen2016} exist
that use different metrics to evaluate trustworthiness.
Network-based indication schemes \cite{Karkazis2012, Chen2011} are trust
management protocols that use metrics in the same way as Antilizer.
The scheme presented in \cite{Karkazis2012} can only detect a single attack
(selective forwarding) due to its use of single metric (forwarding indication).
Antilizer uses more metrics and can detect a wider range of attacks.
The scheme presented in \cite{Chen2011} detects a wider range of
attacks.
Antilizer has much better performance than \cite{Chen2011}, it has a
higher rate of detection of malicious nodes and is better able to mitigate the
effects of attacks by maintaining a higher packet-delivery ratio across the
network.

We do not compare ourselves to the schemes presented in \cite{Bao2012,
Chen2016} because they use metrics collected from neighbour nodes with the
assumption that the information that they receive is trustworthy.
This assumption is dangerous because a malicious node can provide false
information and subvert both schemes.
Antilizer uses network metrics that it has collected itself using overhearing
of its local neighbours without using the content of the communication.
This approach prevents malicious nodes from spreading false information in
the network.
In \cite{Samian2017}, the authors propose a scheme to filter false
recommendation created by dishonest nodes.
This scheme is limited to only detect attacks that falsify information, and
can not detect attacks that subvert a network in other ways, such as not
following a protocol. 

\textbf{Automated response systems.}
There are very few WSN security schemes that detect attacks and prevent
the disruption from the attack.
Antilizer does both.
Examples of systems that do provide an automated response upon detecting
malicious activity are \cite{Sultana2014, Lu2015}.
Neither scheme is trust based, in contrast to Antilizer.

Kinesis \cite{Sultana2014} uses policy specification to select an appropriate
response to a detected attack.
The selected response action is based on a voting scheme which requires
interaction and message exchange between nodes in the same neighbourhood.
The final decision to revoke or reprogram a node is made only by the base
station.
This approach can be slow to detect an attack, and can itself be compromised
by altering communication with the base station or using dishonest nodes.  

The work in \cite{Lu2015} addresses queue-based protocols.
It monitors the queues to detect metric deviation and discover malicious behaviour.
This approach cannot be adapted to distance vector routing protocols such
as RPL without the addition of queues.
Antilizer monitors a larger range of metrics than just message queue lengths,
and is therefore responsive to a greater variety of attack types or behavioural
changes.

\section{System Model} \label{sec:model}

Before we present the security architecture of Antilizer, we define its system model and
our assumptions.

\textbf{Network Model.}
We consider a WSN that has multiple devices $\mathcal{N}=\mathcal{S}\cup\mathcal{R}$ 
communicating  in a multi-hop fashion, where $\mathcal{S}$ is the set of all sensor nodes
generating and relaying data packets, and $\mathcal{R}$ is the set of all roots/base
stations collecting data packets from the network.
In this paper we only use one base station $\mathcal{R}$.
The network operates over a finite-horizon period consisting of discrete time slots
$t \in \{1, 2, \ldots,  t_f\}$, $t_f<\infty$.
We define $\mathcal{N}_{x}(t) \subseteq \mathcal{N}$ to be the set of one-hop neighbours
that node $x \in \mathcal{N}$ can communicate with during time slot $t$.
The network is modelled as a time-varying weighted graph $G(\mathcal{N},\mathcal{L})$. $\mathcal{L}$ is the set of all possible wireless links for the node pairs
$x,y \in \mathcal{N}$.
The entry $(x,y) \in \mathcal{L}$ is the communication link between the source
node $x$ and the destination node $y$.

\textbf{Security Model.}
We consider the base station as trusted with a secure mechanism of disseminating updates
(use of cryptographic keys or secure channels) to the network.
The base station makes the final decision on whether to initiate a request for revoking or
reprogramming potentially malicious nodes.
Even with secure communication from the base station, any individual sensor may become
untrusted and potentially malicious over time.
We assume that each node trusts itself.
The majority of nodes in any neighbourhood are non-malicious.
The existence of a majority of non-malicious nodes ensures the existence of at least one
alternative, non-malicious, route to the base station.

\textbf{Threat Model.}
Assuming the OSI network architecture model as it is applied to WSN, we address attacks
specific to the network layer.
This layer provides data routing for network communication.
Attacks at this layer aim to reduce or delay the flow of sensor data to the base station \cite{Tomic2017}.
An attacker disrupts the flow of data by undertaking one or more of the following malicious
activities:
\begin{itemize}
\item \textit{Falsify information} - The attacker intentionally sends false information
to other nodes to affect their routing decisions.
Examples include the sinkhole (node advertises the false rank) and sybil attacks (node presents
multiple identities in the network).
Both attacks result in the compromise of transmission routes.
\item \textit{Fail to transmit} - The attacker does not obey routing decisions and fails to 
act as a router for its neighbours.
This attack degrades successful data reception by the base station.
Examples range from the most severe case of failing to forward any data packets in the case
of blackhole attack, to the selective forwarding attack where data of only a small set of 
neighbours is forwarded.
\item \textit{Data Injection} - The attacker can inject false information,
replay overheard information, or flood the network with a high rate of
communication.
This attack forces the nodes to waste energy due to increased message reception
and interference in the network.
The result reduces the ability of the network to carry useful data to the base station.
\end{itemize}

While we aim to cover the major network level attacks for WSN we realize that this
categorization is not exhaustive nor can be given the very nature of security.
The approach that we present is engineered to be extensible to new attacks and
can be updated when new security issues arise over time.
We assume that the percentage of the network affected by an attack
is dependant upon the percentage of the network nodes that have become malicious.
The percentage of malicious nodes varies from non-intrusive (1\% to 3\%) to
intrusive (up to 10\%). 

\section{Overview of Antilizer}
 
\begin{figure}
\centering
\includegraphics[width=0.95\linewidth]{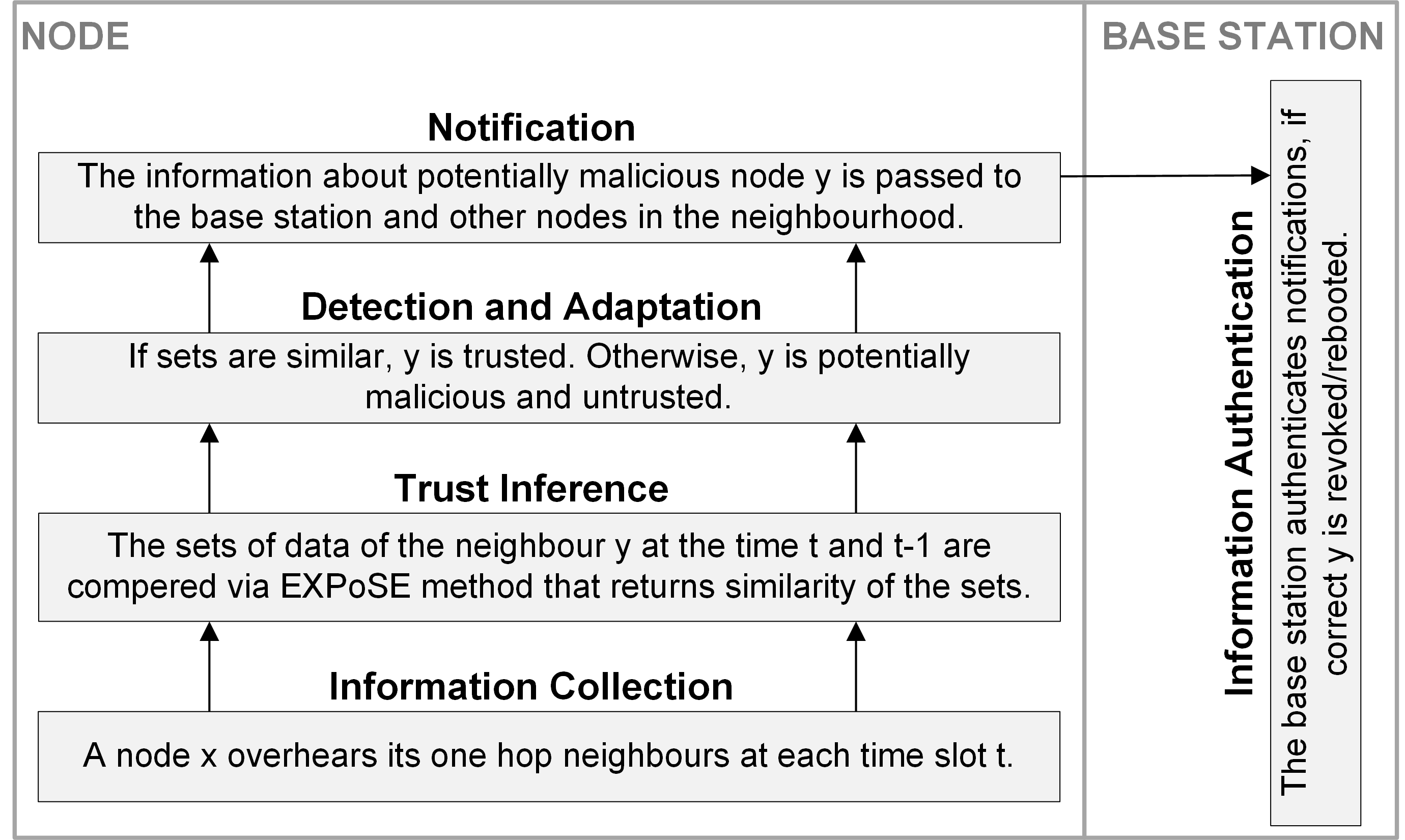}
\caption{The Antilizer architecture.\vspace{-5mm}}
\label{fig:architecture}
\end{figure}

In this section we introduce Antilizer, a novel self-healing security solution for WSNs.
Each node \emph{collects information} to \emph{create a trust model} of their
neighbours.
If a node \emph{detects} any malicious activity from one of its neighbours, it
changes its trust of its neighbour, and \emph{adapts} its communication
decisions based on that trust.
When a malicious node is detected, it \emph{notifies the base station} with
the ID of the suspected node.
The base station then \emph{authenticates} this network security information.
Antilizer arranges these tasks into five modules, each briefly explained below
and depicted in Fig.~\ref{fig:architecture}.

\textbf{Information Collection.}
Each node uses its own radio to overhear the communication and collect information
such as a number of transmissions, number of receptions, for each neighbour.
This information is recorded every time slot $t$, as mentioned in the network model.
These network metrics are used to build a profile of each neighbour in reception range.
As network metrics are overheard without using the content of the communication, 
the node can not be affected by dishonest information shared by its neighbours.

\textbf{Trust Inference.}
The collected information is used as an input to the Expected Similarity Estimation
method (EXPoSE) \cite{Schneider2016}.
In Antilizer we adapt this method to the resource constraints of sensor
nodes by the use of approximations to reduce the storage complexity.
The node uses the expected similarity of collected information sets
over time to infer a trust value for each neighbour.
A neighbour is considered trusted if the two consecutive sets are similar, or the
difference between two consecutive sets is small.
If two consecutive sets are not similar, their difference will be large, and
the associated node will be considered malicious.

\textbf{Detection and Adaptation.}
Large changes in the collected information of a neighbour over time indicates
malicious behaviour.
This causes the detecting node to reduce its trustworthiness towards that neighbour.
Accordingly, the node autonomously adapts its decisions regarding collaboration
with that less trusted neighbour.
We illustrate this via an example of data routing.
The node maps neighbour trust values to a weight used in the routing algorithm's
objective function.
Less trusted neighbours are punished and the data is routed around those
neighbours.

\textbf{Notification.}
Trust models are built in a completely distributed way, each node has a unique
view of the network.
To strengthen an individual node's judgement and prevent the disruption of the
network, nodes need to inform the base station of potential malicious nodes,
and their neighbours of changes to the network caused by suspected malicious
behaviour.
Antilizer uses a simple but smart notification scheme where information is
spread by mobile agents.

Consider a routing example.
When a node detects an anomalous behaviour in its current parent neighbour
(the node to which it routes data), the neighbour will be punished by low
trust value.
The node will then change its parent from that neighbour to another with a
higher trust.
After it changes parents, it triggers an ANT that travels along the route
from the node to the base station carrying the ID of the potentially malicious
parent.
The ANT also informs the neighbours of the new parent that its metrics
will be changing in the short term due to a potential security issue.
In this example, ANTs bound the damage caused by an attack and stop the
spread of distrust in the network.

\textbf{Information Authentication.}
The information that is passed by an ANTs to the base station is used to
determine an appropriate reaction to a potentially malicious node, such as
to revoke or reboot that node.
To verify the authenticity of the information, we use a filtering mechanism.
If the information is verified as correct, the malicious node will be punished.
Otherwise, the information will be discarded and considered as malicious, and as an
indication of further malicious activity.

A detailed description of individual modules is given in the further sections.

\section{A Self-Referenced Trust Model to Detect Potentially Malicious Nodes}
\label{sec:trust_routing}

This section describes the Antilizer trust model in detail.
First, we give an explanation of the network metrics used for neighbourhood
surveillance and our method of collecting information through overhearing.
Then, we present in detail how we use the network metrics with the EXPoSE method to
infer trust.

\subsection{Neighbourhood Surveillance with Network Overhearing}
\label{subsec:Get_Network_Metrics}

WSN nodes make communication decisions regarding collaboration with a neighbour
node without taking into account that the neighbour might become malicious and
violate the underlying protocol rules.
For example, a malicious neighbour can alter data or falsify shared information. 
To address this problem, we introduce trustworthiness as an additional metric to be 
used by a node when making these decisions.

\begin{figure}
\centering
\includegraphics[width=0.8\linewidth]{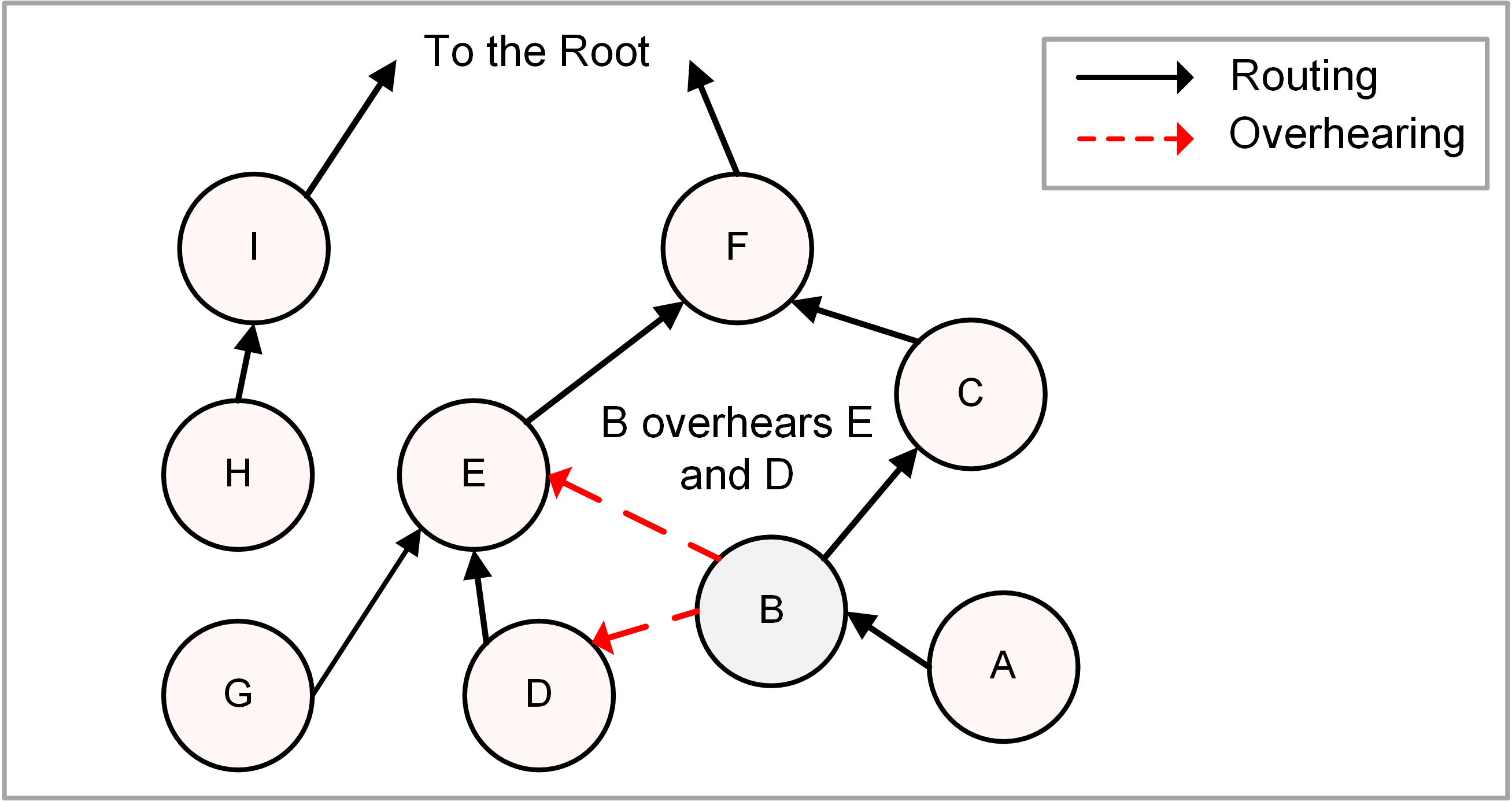}
\caption{The overhearing phenomenon.\vspace{-5mm}}
\label{fig:limitations}
\end{figure}

\textbf{Network Overhearing.}
Nodes collect network behaviour information for all of their one-hop neighbours. 
Information collection is done by overhearing radio packet transmissions in their
reception range even if they are not intended recipients \cite{Paris2013}.
The use of overheard information does not guarantee the capture of all one-hop
neighbour information.
Fig.~\ref{fig:limitations} illustrates where node B can overhear packets sent
from D to E; however, it cannot hear any packets that E might have received from G.
Additionally, leaf nodes such as G, do not perform any forwarding tasks; thus not
all of their metric can be measured.
Despite these limitations, the information collected using network overhearing
is sufficient to describe nodes' behaviour and infer their trust values.

\textbf{Network Metrics.}
The use of overhearing overcomes the problem of using potentially dishonest
information.
Node $x$ trusts only itself and the information that it can overhear without
using the content of the communication.
This information is stored as a set of network metrics.
Table~\ref{tab:network_metrics} lists the metrics collected by node $x$ for each
of its one-hop neighbours $y\in \mathcal{N}_x(t)$ within a given time slot.
These are added to a network metric vector $\mathbf{v}_{x,y}(t)\in \mathbb{R}^d$
at every time slot $t$, where $d=3$ is the number of metrics.

\begin{table}
\renewcommand{\arraystretch}{0.9}
\caption{Network metrics retrieved via overhearing}
\label{tab:network_metrics}
\centering
\begin{tabularx}{0.48\textwidth}{l X}
\toprule
\textbf{Metrics} & \textbf{Description} \\
\midrule
    \textit{Tx}
    &the number of packets the node $y$ transmitted to its neighbours\\
    \textit{Rx/Tx}
    &the ratio of the number of received and the number of transmitted packets at node $y$ (forwarding indication)\\
    \textit{Rank}
    &the average rank of the node $y$\\
\bottomrule \vspace{-5mm}
\end{tabularx}
\end{table}

\subsection{Temporal Similarity of Network Metrics}
\label{subsec:Trust}

The network metrics collected via overhearing change over time.
They do not follow a specific distribution.
The use of a parametric solution, such as a Gaussian distribution, reduces the 
reliability of the results obtained.
Instead of assuming a certain distribution, we exploit a state-of-the-art non-parametric
technique called the Expected Similarity Estimation method (EXPoSE) \cite{Schneider2016}.
We choose ExPoSE because of its proven accuracy and its ability to be altered
to work on low-power devices such as sensor nodes.
We use EXPoSE with a set of approximations to reduce its storage requirements.
Without these approximations EXPoSe would be unusable on resource-restricted sensor
nodes.

\textbf{Network Metrics Similarity.} 
In the EXPoSE method, the network metrics are combined into a vector
$\mathbf{v}_{x,y}(t)$ for each time slot $t$.
The vectors are then mapped into Hilbert space $\mathcal{H}$.
The mapping is done using the function $\phi$ at every time slot $t$.
The change between two consecutive vectors is described through the expected
similarity measure defined below.

\begin{Definition}[Expected similarity] \label{def:expected_similarity}
Given a node $x$ and $y\in \mathcal{N}_x(t)$, the expected similarity between the vector
$\mathbf{v}_{x,y}(t)$ at time $t$ and its previous version at time 
$(t-i)$, $\mathbf{v}_{x,y}(t-i)$, for $i=1,2,...,n$ with $n$ being the number of past
vectors, is defined as 
\begin{eqnarray} \label{eq:expeted_similarity}
  \eta(\mathbf{v}_{x,y}(t))
  &=&\sum_{i=1}^n k(\mathbf{v}_{x,y}(t),\mathbf{v}_{x,y}(t-i)) \nonumber\\
  &=&\sum_{i=1}^n\langle \phi(\mathbf{v}_{x,y}(t)),\phi(\mathbf{v}_{x,y}(t-i))\rangle.
\end{eqnarray}
The kernel function $k(\cdot,\cdot)$ computes the inner product of two vectors in 
$\mathbb{R}^d$ after mapping them into higher dimensional Hilbert space 
$\mathcal{H}\in \mathbf{R}^{d'} (d' \geq d$) through a mapping function $\phi(\cdot)$.
The term $\langle\cdot,\cdot\rangle$ denotes the inner product of the two vectors.
For more information regarding the definition of the kernel function, we refer readers
to \cite{hsu2003practical}.
\end{Definition}

Although the kernel function $k(\cdot,\cdot)$ allows the computation of the inner
product without 'visiting' the high dimensional Hilbert space $\mathcal{H}$, it does
not support the computation of the inner product in an incremental manner.
The computation of the inner product in a incremental way is important to the
implementation of this method on resource-constrained sensor nodes.
Incremental updates allow the accurate capture of the data stream dynamics of
$\mathbf{v}_{x,y}(t)$ in way that requires minimal computation and storage.
This is achieved because $\eta(\mathbf{v}_{x,y}(t))$ does not have to be recomputed
at every time slot $t$.
We realise incremental updates through the use of a kernel approximation (KEA)
vector which we defined below.

\textbf{KEA Vector.}
By using the KEA vector, $\mathbf{\si{\micro}}_{x,y}(t)$, Eq.~\eqref{eq:expeted_similarity}
becomes:
\begin{eqnarray} \label{eq:kernel_embedding_approx}
\eta(\mathbf{v}_{x,y}(t))&=&\langle \phi(\mathbf{v}_{x,y}(t)),\frac{1}{n}\sum_{i=1}^{n}\phi(\mathbf{v}_{x,y}(t-i))\rangle \nonumber \\
&=&\langle \phi(\mathbf{v}_{x,y}(t)),\mathbf{\si{\micro}}_{x,y}(t)\rangle
\end{eqnarray}
where $\mathbf{\si{\micro}}_{x,y}(t)$ can be updated in an incremental manner as:
\begin{equation} \label{eq:ewma}
	{\si{\micro}}_{x,y}(t)=\gamma\phi(\mathbf{v}_{x,y}(t-1))+(1-\gamma){\si{\micro}}_{x,y}(t-1).
\end{equation}
The term $\gamma\in [0,1]$ denotes an automatic decay factor that controls the
speed at which the trust metric will change to the occurrence of new information.
A larger $\gamma$ results in the faster decay of past information.

The KEA vector ${\si{\micro}}_{x,y}(t)$ is computed using the overheard network
metrics collected in previous time slots, $\mathbf{v}_{x,y}(t')$, where $t'<t$.
The metrics used for the update of ${\si{\micro}}_{x,y}(t)$ need to come from a
trusted, non-malicious node.
The network metrics generated from potentially malicious nodes are detectable when
they first occur because the difference between the current and past metrics will
be large.
As time continues $\mathbf{\si{\micro}}_{x,y}(t)$ will change to incorporate the
new network metrics.


To prevent the adaptation of $\mathbf{\si{\micro}}_{x,y}(t)$ to a malicious
node we define a parameter $\alpha$ to determine which overheard network
metrics get included in the KEA vector. 
When $\eta(\mathbf{v}_{x,y}(t))>\alpha$ the network metrics for that time slot
$t$ will not be used to update the KEA vector, 
(i.e. ${\si{\micro}}_{x,y}(t)={\si{\micro}}_{x,y}(t-1)$), as the new behaviour 
is likely to be malicious.
According to our extensive simulations in Sec.~\ref{subsec:results}, $\alpha \in [0.7,0.9]$
ensures the detection of more than $98.6$\% anomalies with low rate of false positives 
($\sim3.6$\%) (see Fig.~\ref{fig:det_reliab}).

\begin{figure}
\centering
\includegraphics[width=0.95\linewidth]{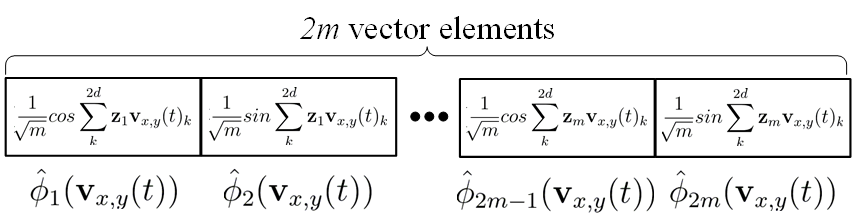}
\caption{Approximation of the mapping function $\hat{\phi}(\mathbf{v}_{x,y}(t))$
results in $2m$ elements.\vspace{-5mm}}
\label{fig:phi_approx_example}
\end{figure}

\textbf{Mapping Function Approximation.}
We use the radial basis function (RBF) \cite{hsu2003practical} as the kernel
function to compute KEA.
RBF has been widely used in many machine learning applications, including non-parametric
regression, clustering and neural networks.
We can not directly use RBF because it transforms data vectors into an
\emph{infinite} dimensional space (i.e. $\phi(\mathbf{v}_{x,y}(t))\in\mathbb{R}^\infty$
and $\mathbf{\si{\micro}}_{x,y}(t)\in\mathbb{R}^\infty$) whose storage would
exceed that of a sensor node.
We fix this problem with the use of a Monte Carlo approximation of the RBF
mapping function.
An inverse Fourier transform is used to approximate the Gaussian RBF kernel \cite{rahimi2008random}.
The approximation function $\hat{\phi}(\mathbf{v})$ is given by the Euler equation:
\begin{equation} \label{eq:phi_approximation}
\hat{\phi}(\mathbf{v}_{x,y}(t))=\frac{1}{\sqrt{m}}exp(i\mathbf{Z\mathbf{v}_{x,y}(t)})
\end{equation}
where $i$ denotes the imaginary unit ($i^2 = -1$), $\mathbf{Z}$ is a $m\times
d$ matrix where each element $\mathbf{Z_{i,j}}\sim \mathcal{N}(0,1/\sigma^2)$,
and $m$ represents the number of Monte Carlo samples.
The operation in Eq.~\ref{eq:phi_approximation} is depicted in
Fig.~\ref{fig:phi_approx_example}.
The Euler equation results in $2m$ elements where the odd elements store the real
parts and the even elements store the imaginary parts of the complex number.
The expected similarity $\eta(\mathbf{v}_{x,y}(t))$ in 
Eq.~\ref{eq:kernel_embedding_approx} then becomes:
\begin{equation}\label{eq:mapping_approximation}
\eta(\mathbf{v}_{x,y}(t))\approx\langle \hat{\phi}(\mathbf{v}_{x,y}(t)),\frac{1}{n}\sum_{i=1}^{n}\hat{\phi}(\mathbf{v}_{x,y}(t-i))\rangle
\end{equation}
where $1/n\sum_{i=1}^{n}\hat{\phi}(\mathbf{v}_{x,y}(t-i))$ is computed in an
incremental manner with Eq.~\eqref{eq:ewma}.

The expected similarity computation method which uses Monte Carlo
approximation is given in Alg.~\ref{alg:ex_sim}.
The computational complexity of Alg.~\ref{alg:ex_sim} reduces from $O(mnd)$
to $O(m\log d)$ for all $t$, where $m$ denotes the number of Monte Carlo samples,
$n$ denotes the number of past vectors, and $d$ denotes the number of network
metrics obtained at every time $t$ \cite{Schneider2016}.
This reduction in complexity is a result of $\phi(\mathbf{v}_{x,y}(\cdot))$ not
having to be recomputed in each time slot.
Note that, $\eta(\mathbf{v}_{x,y}(t))\in [0,1]$ is normalized by the $1/\sqrt{m}$
in Eq.~\eqref{eq:phi_approximation}.
\begin{algorithm}[tp] 
\small
\DontPrintSemicolon
\LinesNumbered
\SetKwInOut{Input}{Input}
\SetKwInOut{Output}{Output}
\Input{
	$\mathbf{v}_{x,y}(t)$: the network metrics of $y$ observed from $x$ at time $t$\\
    $\mathbf{\si{\micro}}_{x,y}(t)$: the KEA vector of $y$ observed from $x$ at time $t$\\
	$\sigma^2$: the standard deviation for Monte Carlo sampling\\
    $\mathbf{z}$: the vector of $m$ Gaussian samples from $\mathcal{Z}~(0,1/\sigma^2)$\\
    $\alpha$, $\gamma$: the parameters to control KEA vector adaptation\\
}
\Output{$\eta(\mathbf{v}_{x,y}(t))$: the expected similarity of the network metrics of $y$ observed from $x$ at time $t$}
$\hat{\phi}(\mathbf{v}_{x,y}(t))\leftarrow 0, \mathbf{\si{\micro}}_{x,y}(1)\leftarrow \hat{\phi}(\mathbf{v}_{x,y}(1)), \eta(\mathbf{v}_{x,y}(t))\leftarrow 0$\\
/* STEP 1: mapping function approximation */\\
\For{$j\Leftarrow 1,2...,m$}{
	$\hat{\phi}_j(\mathbf{v}_{x,y}(t))\leftarrow \frac{1}{\sqrt[]{m}}exp(i\mathbf{z}\cdot\mathbf{v}_{x,y}(t))$ \\
}
/* STEP 2: compute expected similarity */\\
$\eta(\mathbf{v}_{x,y}(t))\leftarrow \langle\phi(\mathbf{v}_{x,y}(t)),\mathbf{\si{\micro}}_{x,y}(t)\rangle$ \\
/* STEP 3: update KEA vector for time $t+1$ */\\
\If{$k>1$ and $\eta(\mathbf{v}_{x,y}(t))<\alpha$}{
	${\si{\micro}}_{x,y}(t)=\gamma\phi(\mathbf{v}_{x,y}(t-1))+(1-\gamma){\si{\micro}}_{x,y}(t-1)$
}
\Return{$\eta(\mathbf{v}_{x,y}(t))$}
\caption{Expected similarity $\eta(\mathbf{v}_{x,y}(t))$ computation}\label{alg:ex_sim}
\end{algorithm}

In the next section we show how the similarity measure is mapped to a trust value and
used to enhance the objective function of the routing algorithm.

\section{Application of the Self-Referenced Trust Model to Data Routing}

In this section we describe how the trust, computed from the similarity measure of
network metrics, is used in the routing algorithm to affect routing decisions.
The routing is an example of node collaboration where communication decisions
can be based on the notion of trust.
First, we discuss the class of routing protocols that can benefit from our scheme.
Then, we show how our trust metric can be used by a routing algorithm objective
function to avoid areas affected by a network level attack.

\subsection{Distance Vector Routing Protocols: RPL as an Example} \label{subsec:RPL}

Antilizer can be applied to any distance-vector routing protocol where a distance
measure (e.g. hop count, or respective link qualities) is used to determine the
best packet forwarding route.
In this paper we use RPL (Routing Protocol for Low-Power and Lossy Networks)
\cite{rfc6550}. 
RPL is a standardized IPv6-based multi-hop routing solution widely used in WSNs.
It is an appropriate protocol for the evaluation of Antilizer because it requires
an objective function where the trustworthiness can be added as an additional metric.
It is important to mention that Antilizer affects only the objective function;
therefore, it can be easily applied to any other routing protocol based on a
distance-based objective function.

An objective function defines how a node translate one or more network metrics and
constraints into a $\rm Rank$ value.
$\rm Rank$ is used to determine the best neighbour to forward data to the base station.
The $\rm Rank$ between the node $x$ and its neighbours $y\in \mathcal{N}_x(t)$
is given by
\begin{equation}
\label{eq:rplRank}
\rm Rank_{x}(t) = \left\{
   \begin{array}{l l}
    \underset {y \in \mathcal{N}_{x}(t)} {\min}( p_{x,y}(t) + \rm Rank_{y}(t))& \quad x\notin\mathcal{R} \vspace{0.2em}\\
    {\rm RootRank}_x & \quad x\in\mathcal{R}
   \end{array} \right.
\end{equation}
where $p_{x,y}(t)>0$ denotes the penalty of using the link $(x,y)$ at time slot $t$,
and ${\rm RootRank}_x \geq 0$ is the smallest $\rm Rank$ value in the routing tree.
The smallest $\rm Rank$ value in a correctly operating network belongs to the root $x$.

The $mrhof$ objective function is used in the Contiki implementation of RPL.
It defines $p_{x,y}(t)$ as a moving average function that uses the expected number
of transmissions ($\ETX$) as the routing measure \cite{Contiki}:
\begin{equation} \label{eq:link_penalty}
p_{x,y}(t)=\textsf{ALPHA}p_{x,y}(t-1)+(1-\textsf{ALPHA})\ETX_{x,y}(t)
\end{equation}
where {$\textsf{ALPHA}=\{0.15,0.3\}$}. 
In our work we add the trustworthiness of individual nodes to the routing measure.
This enhanced objective function is given next.

\subsection{Trust-based RPL Objective Function}
\label{subsec:Trust_OF}

The expected similarity $\eta(\mathbf{v}_{x,y}(t))$ of the network
metric vector $\mathbf{v}_{x,y}(t)$ at time $t$ is included in the $\rm Rank$ computation in
Eq.~\eqref{eq:rplRank} through the link penalty function in Eq.~\eqref{eq:link_penalty}.
It is included as subjective 'penalty', or weighting function, called the
\emph{Subjective Trust}.

\begin{Definition}[Subjective Trust]\label{def:subjective_trust}
Given a node $x$ and its one-hop neighbour $y\in \mathcal{N}_x(t)$, the subjective
trust $\tau_{x,y}(t)$ is the penalty that $x$ gives to $y$ at time $t$.
Its value is defined through the hyperbolic function:
\begin{equation}\label{eq:hyperbolic_function}
	\tau_{x,y}(t)=-csch(k\eta(\mathbf{v}_{x,y}(t))-k)+csch(-k)+1
\end{equation}
where $csch(\cdot)$ denotes the hyperbolic cosecant function, and $k$ is a parameter
used to control the detection sensitivity of our solution. 
\end{Definition}

\begin{figure}
\centering
\includegraphics[width=0.95\linewidth]{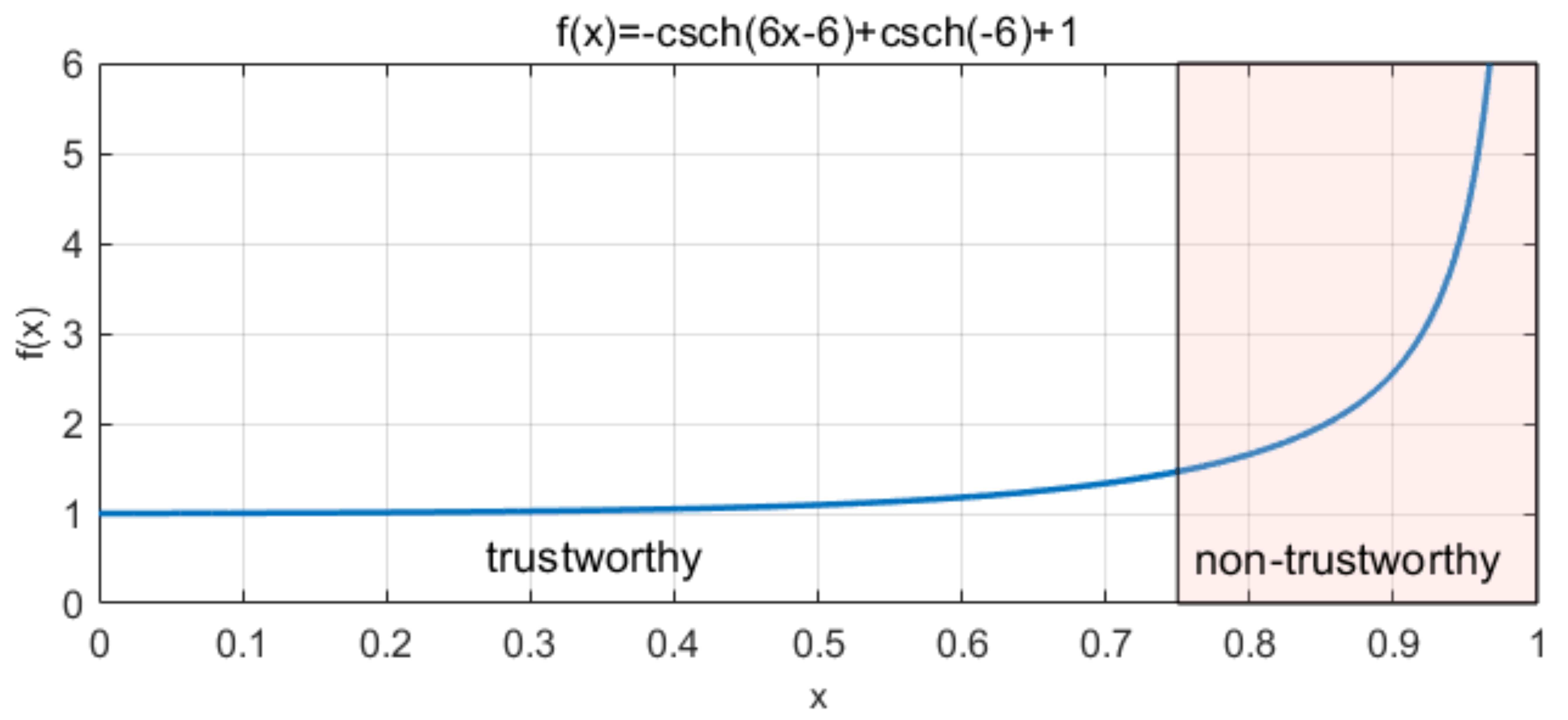}
\caption{Hyperbolic cosecant function with $k=6$.\vspace{-3mm}}
\label{fig:hyperbolic_function}
\end{figure}

This function provides the continuous transient where $\tau_{x,y}(t)\approx 1$ for all
$\eta(\mathbf{v}_{x,y}(t))\leq\alpha$.
For $\eta(\mathbf{v}_{x,y}(t))> \alpha$, the value of $\tau_{x,y}(t)$
grows exponentially.
By adjusting $\alpha$ and $k$, the sensitivity of the scheme can be controlled.
Fig.~\ref{fig:hyperbolic_function} illustrates the hyperbolic function where
$\alpha=0.75$ and $k=6$.
Function $f(x)$ takes as an input the expected similarity $\eta(\mathbf{v}_{x,y}(t))$ obtained in Alg.~\ref{alg:ex_sim} and it returns the subjective trust $\tau_{x,y}(t)$.
%

As can be observed from Def.~\ref{def:subjective_trust}, $\tau_{x,y}(t)$ is bounded
between $1$ and $\infty$, i.e. $\tau_{x,y}(t) \in [1,\infty]$.
$\tau_{x,y}(t)=1$ indicates that node $y$ is $100$\% trustworthy from the perspective of
node $x$.
Large values of $\tau_{x,y}(t)$ indicate the reduction of trustworthiness towards the node $y$.
This measure of the trustworthiness of individual nodes is then used as a
weighting function in the link penalty function $p_{x,y}(t)$ in Eq.~\eqref{eq:link_penalty}:
\begin{equation}
\small
\label{eq:penalty_new}
\hat{p}_{x,y}(t) = \textsf{ALPHA}\hat{p}_{x,y}(t-1)+(1-\textsf{ALPHA})\tau_{x,y}(t)\ETX_{x,y}(t).
\end{equation}

The addition of trustworthiness directly affects a nodes routing decisions and
ensures that it avoids malicious nodes so that the flow of data is not
obstructed during an attack.

\begin{Remark}[Zero performance penalty in no attack scenario]
The hyperbolic function illustrated in Fig.~\ref{fig:hyperbolic_function} ensures a high
penalty when potentially malicious behaviour is detected.
It also guarantees optimal performance (i.e. as when default objective function of RPL is
used) in no attack scenario, that is, $\hat{p}_{x,y}(t)\approx p_{x,y}(t)$.
\end{Remark}

Next, we present an agent-based notification scheme to address the limitations of the
proposed trust-based scheme.

\section{Mobile agent-based notification scheme to stop the disruption of the network}
\label{sec:Ants}

While the self-referenced trust model proposed in Sec.~\ref{sec:trust_routing} is able
to detect potentially malicious change and adapt node's communication decisions such 
that the affected area is avoided, it has its own limitations.
The scheme can not completely prevent the disruption to the network because the
network will eventually require reconfiguration to handle the malicious node,
and the adaptation of one node may be interpreted as malicious behaviour by another.
To address these issues, we present a notification scheme that uses mobile agents called
ANTs (Antilizer Notification ticket) to inform the base station of suspected malicious
behaviour, to bound the damage of the attack to a specific area, and prevent the spread
of distrust in the network.

First, we introduce the features of ANTs and their working principle.
We show how these are applied to the routing example.
Finally, we discuss how the information carried by ANTs can be authenticated at the
base station.

\subsection{Overview of the ANT and ANT's Features}
\label{subsec:Ants_Overview}

When node $x$ detects a change in network behaviour of its parent $y$, it reduces its
trustworthiness.
In the example of routing this leads to $x$ changing the routing path (i.e. $x$ chooses
a node with the lowest rank in the neighbourhood).
At the same time, $x$ creates an ANT which does the following:
%
\begin{enumerate}
\item It migrates from the node $x$ to its new parent $y'$ while carrying the ID of the
potentially malicious node (i.e. $x$'s old parent $y$). 
\item After it successfully migrates to $y'$, it triggers a broadcast message to all 
one-hop neighbours of $y'$, $\mathcal{N}_{y'}(t)$, to notify them of potential malicious
activity in the neighbourhood, and to prevent the spread of distrust caused by
network variations.
\item The ANT then moves to the next hop along the established route and repeats this
process until it reaches the base-station.
\item The ANT delivers the ID of the potentially malicious node to the base-station which
decides if node $y$ should be revoked/reprogrammed (may involve human interaction).
\end{enumerate}
%

Next, we further discuss the features of this notification scheme, as well as the
introduced overhead.

\textbf{Protection at Low Overheads.} 
As described before, an ANT does two operations: it travels to the base-station
along the RPL tree, and it broadcasts a one-hop message at each hop in the route.
Given a node which has a route to the base station of $z$-hops, the number of extra
messages introduced by each ANT is $2z+1$.
ANT messages require no more than one data packet of less than $160$~bytes.
Therefore, the operations done by ANTs are very lightweight and with low operational
overhead as shown in Sec.~\ref{subsec:results} (less than $1$\%).

\textbf{Guarantees for No Attack Scenarios.}
ANTs will be created only by nodes that detect a change in network behaviour of
their parent.
No ANT will be spawned when no attacks occur, and no extra communication
is required.
Our experimental results in Sec.~\ref{subsec:results} show that the percentage of
false positive detections is less then 3.4\% for $\alpha \in [0.7,0.9]$,
and there is a zero performance penalty when Antilizer is run no attack scenario.

\begin{figure}
\centering
\includegraphics[width=0.7\linewidth]{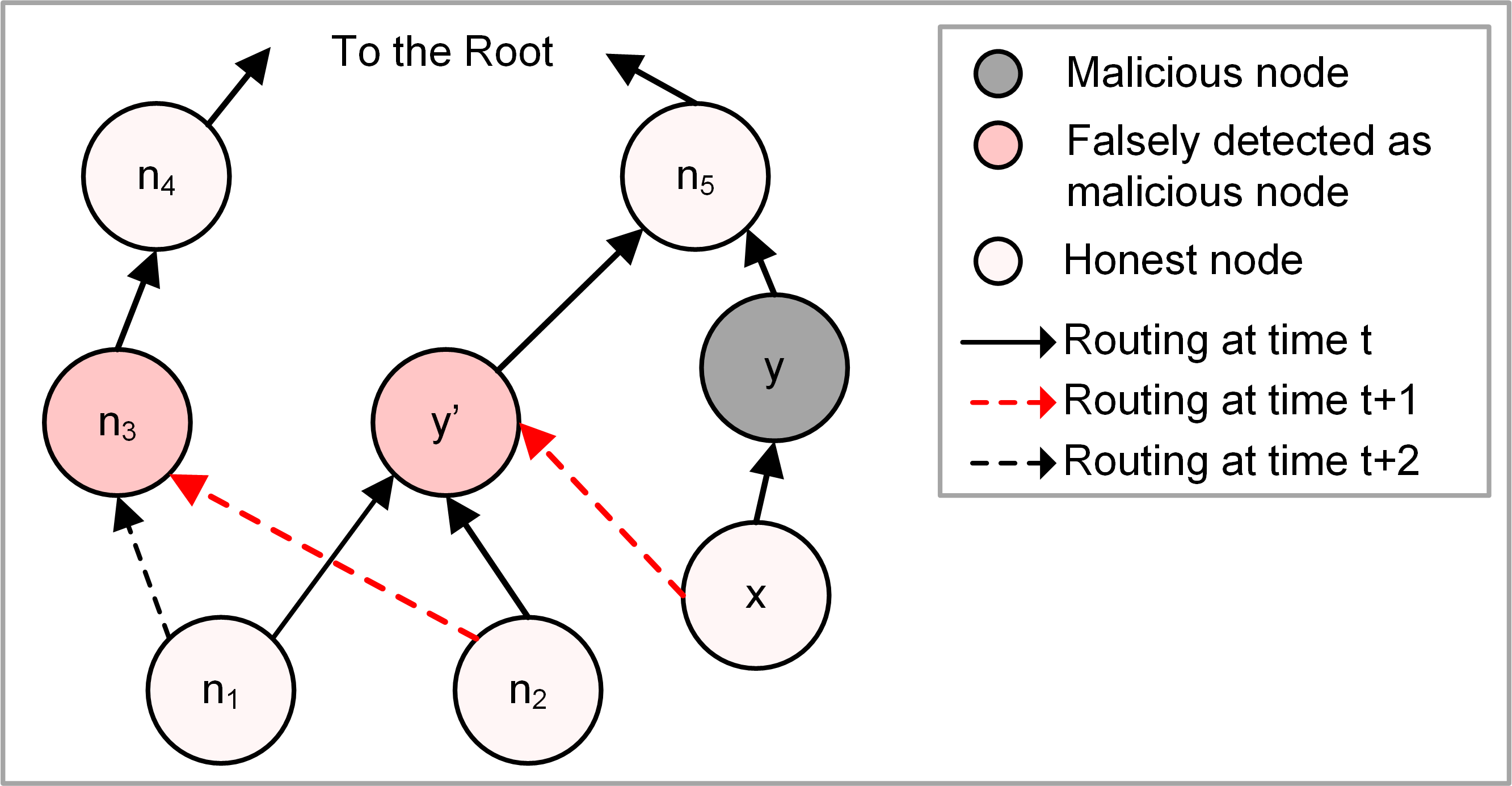}
\caption{An illustration of a attacker's neighbourhood (node $y$ is the attacker
that has been chosen by node $x$ as its parent, $x$ and $n_1$-$n_5$ are honest
nodes).\vspace{-4mm}}
\label{fig:terminal}
\end{figure}

\textbf{Distinguishes Genuine Change from Malicious Behaviour.}
When node $x$ detects a malicious behaviour in $y$ it changes to an alternative
route through $y'$ (which has the lowest rank in the neighbourhood) and sends its
ANT towards the base station (see Fig.~\ref{fig:terminal}).
This change increases the traffic of $y'$.
From the perspective of its neighbours $n_1$ and $n_2$, $y'$ is likely to be seen as
malicious due to its increasing $\rm T_x$ and $\rm R_x$.
As a result, node $y'$ is penalized by this false positive detection, and $n_1$ and
$n_2$ switch their routes to $n_3$, which further spreads distrust.
If this spread is left unchecked, nodes will run out of 'safe' routes and RPL
will be unable to converge to a stable routing tree.

In our scheme, an ANT created by $x$ triggers a broadcast message at its new RPL-tree
parent $y'\in\mathcal{N}_x(t)$.
The ANT informs all of the one-hop neighbours of $\mathcal{N}_{y'}(t)$ (e.g. nodes $n_1$ 
and $n_2$) of the change at $x$.
Now instead of flagging $y'$ as potentially malicious, nodes allow a certain
period of time (a refractory period) to adapt to the new network behaviour caused by
the change of routes of $x$. 
ANTs reduce the number of false positive detections which improves Antilizer performance
and bounds the damage of an attack to a specific area.


To validate the authenticity of the information provided by ANTs at the base station,
we use a filtering mechanism described next.

\subsection{Credibility of Information carried by ANTs}
\label{subsec:Ants_Credential}

Here we discuss the potential drawbacks of our notification scheme and
scenarios when it can be compromised.

\textbf{Filtering mechanism to authenticate information.}
An attacker can use ANTs to flood a network.
The attacker can create a number of 'falsified' ANTs by switching between its parents.
The base-station would quickly be able to identify this as malicious
activity as ANTs would arrive at high frequency from a single source.
The flooding of ANTs would also not prevent a node from routing away from a malicious
neighbour.
The filtering mechanism at the base station that is run for individual time slot is presented in Alg.~\ref{alg:filter}.
Parameters $\theta_b$ and $\theta_n$ are user defined, depending on the requirements on mechanism sensitivity.
\begin{algorithm}[tp]
\small
\DontPrintSemicolon
\LinesNumbered
\SetKwInOut{Input}{Input}
\SetKwInOut{Output}{Output}
\Input{
    $\rm ANT (a_1, a_2)$: The ANT that carries the ID of the blacklisted node ($a_1$) and the ID of node that reported blacklisting ($a_2$)\\
    $\rm ANT_\text{data} \in \mathrm R^{ixj}$: A matrix which individual entries indicate how many times nodes $\{1,\ldots,i\}$ have been blacklisted by nodes $\{1,\ldots,j\}$ where $i,j \in {\mathcal{S}}$ \\
    $\theta_b, \theta_n$: Parameters that control the filtering mechanism\\
    $n$: A number of $\rm ANT (a_1, a_2)$ that arrived within time slot
}
\Output{$c\in\{\textsf{GD},\textsf{CA},\textsf{FP}\}$}
	$\rm ANT\leftarrow \mathbf{0}$\\
	\For{$l\Leftarrow 1\cdots n$}{
       $\rm ANT(a_1,a_2) \leftarrow \rm ANT_(a_1,a_2)+1$\\
       }
    \For{$k\Leftarrow 1\cdots i$}{
       \If{$|\rm ANT(i,k)|>\theta_b$ /*\rm where $|\rm ANT(i,k)|$ indicates the number of non-zero elements of vector*/}{
          $c\leftarrow\textsf{GA}$ /*Genuine attack, $i$ is malicious*/\\
       }
       \ElseIf{$|\rm ANT(i,k)|<\theta_b$ \text{and} $\rm max(\rm ANT(i,k))\geq\theta_n$}{
          $c\leftarrow\textsf{CA}$ /*Compromised ANT by $k$ for which $\rm ANT(i,k)$ is max*/\\
       }
       \Else{
          $c\leftarrow\textsf{FP}$ /*False positive*/\\
       }
    }
\Return{$c$}
\caption{Filtering mechanism to authenticate information}\label{alg:filter}
\end{algorithm}

\textbf{Encryption to avoid information alteration.}
Our attacker model does not assume the alteration of packet's content.
To mitigate these sorts of attacks there is the need of encrypting the content of
ANT messages.
This is a natural extension of this work.
Encryption alone is usually not sufficient as an attacker could fetch encrypted
packets and replay them. 
Once again, using the filtering mechanism the base station would detect a large number
of ANTs being send from a single node and would be able to identify malicious behaviour.

\textbf{Blacklisting to ensure avoidance.}
Lastly, we discuss the cost of losing some ANTs.
As Antilizer ensures routing around the affected areas, nodes will never re-establish
routes through potentially malicious nodes until these are re-approved by the
base station.
If the base station does not receive any ANT reporting malicious activity, the
area affected by that node will still be avoided.

\section{Implementation}
\label{sec:impl}

We implemented Antilizer in Contiki \cite{Contiki} version 3.1, an open source
operating system for WSNs and IoT.
Contiki provides an IPv6 stack (uIPv6) and the RPL routing protocol.
Our implementation used both.
In this section, we describe key aspects of our implementation.

\textbf{Mac and IPv6.}
For the MAC layer, a CSMA/CA driver was used with default settings in Contiki.
The neighbour table size is set to 50.
The radio duty cycling is disabled during all our experiments.
The maximum transmission attempts to re-send a packet is 5.
In regards to the IPv6 stack, the packet reassembly service is disabled.
The \texttt{UIP\_CONF\_IGNORE\_TTL} is set to zero to ignore the TTL flag in the
packet headers.
The \texttt{HC6} SICSlowpan header compression is used..
The application layer generates UDP packets at a fixed rate of one every four 
seconds.
The size of a data packet is 160 bytes (8 bytes for payload and the rest is used
for IPv6 header). 

\textbf{Attacks.}
Each compromised node $x_c \in \mathcal{N}$ is able to initiate one or more
of the malicious activities described in Sec.~\ref{sec:model}.
We give a specific description of our attack implementation below:
\begin{enumerate} 
\item \textit{Sinkhole attack} - The node $x_c$ advertises falsified rank information
(i.e. it claims that it's a sink, $\rm Rank_{x_c}(t)={\rm RootRank}_x$, $x \in \textrm{R}$.  
\item \textit{Blackhole} - A compromised node $x_c$ advertises falsified rank information
to lure the traffic and then fails to forward any data received from it's neighbours.  
\item \textit{Hello Flood} - The node $x_c$ broadcasts hello packets at a very high rate
to all of its one-hop neighbours $y\in \mathcal{N}_{x_c(t)}$.
\end{enumerate}

\textbf{Routing objective function and metrics retrieval.}
The RPL routing protocol is used with the ETX objective function and default settings.
The \texttt{RPL\_MOP\_NO\_DOWNWARD\_ROUTES} option is enabled since downward routing
is not used in our experiments.
\texttt{DIO\_INTERVAL\_MIN} and \texttt{DIO\_INTERVAL\_DOUBLINGS} broadcast routing
metadata every $512$ to $1024$ms. 

To implement overhearing, we extended the Contiki network link stats module.
The data is stored in a separate neighbour table used only by Antilizer.
The network metrics $\rm T_x$ and $\rm R_x$ are collected by callbacks in
link stats, whereas the public functions are used to retrieve the $\rm Rank$ metric.
The minimum $\rm Rank$ with hysteresis objective function (rpl-mrhof.c) is used as
the base when no malicious behaviour is detected.
The expected similarity value is calculated every $20$ seconds for each node in the
neighbour table, and used as a weight for the calculation of that node's $\rm Rank$.

\textbf{Notification via ANTs.}
The ANTs were implemented using RPL ICMP6 messages and RPL broadcast messages.
When a node changes it's next hop neighbour due to a change in rank weight from
Antilizer, it produces an ANT.
The first hop of the ANT goes from the node to its new upstream next hop and
contains the IPv6 ID of the node considered malicious.
When the ANT is received by the new upstream node, the receiver sends an IPv6
broadcast to all of it's one hop neighbours.
The ANT eventually arrives at the base station where it is verified, and its
data used.

\section{Experimental Evaluation}
\label{sec:evaluation}

In this section, we present the results of an extensive simulation study to evaluate the performance of Antilizer.

\subsection{Experimental Setup}

We performed our evaluation by using the Contiki simulator Cooja.
As a routing protocol, we use the RPL.
We consider networks with $25$, $50$ and $100$ nodes randomly distributed over a $100$m$\times100$m, $200$m$\times200$m, $400$m$\times400$m area, respectively.
Each node has a $\rm T_x$ range of $50$m and periodically sends out data
every 4 seconds with an initial random offset. 

To configure Antilizer we set the number of Monte Carlo samples to $m=200$ with
the standard deviation set to $\sigma^2=0.35$.
We use a hyperbolic cosecant function depicted in Fig.~\ref{fig:hyperbolic_function}.
The parameter $\alpha$ is set to $0.75$ as explained in Sec.~\ref{subsec:results}.
The duration of a time slot is set to $20$ seconds.

We consider a set of attack scenarios in the simulations:
\begin{enumerate}
\item \textit{A single attacker} in the network exploiting one of three attacks
defined in Sec.~\ref{sec:impl}, 
\item \textit{A multi-attacker case} where the attackers belong to the same category
of attacks (i.e. we vary an attack for non-intrusive to an intrusive scenario with 
$10$\% of the nodes being malicious).
\end{enumerate}
Simulations are run for $4$~hours of simulation time with the attack starting time at
$20$~minutes for a $25$~nodes simulation or $40$~minutes for a $50$~nodes and $100$~nodes
simulation.

\subsection{Performance Metrics}

The effectiveness of Antilizer is evaluated based on the following metrics:
\begin{itemize}
\item \textit{End-to-End (E2E) Data Loss} - Ratio between the total number of packets
successfully received by the base station and the number of packets sent by the nodes.
\item \textit{Average End-to-End (E2E) Delay} - Average time needed for a packet to travel
between the source and the base station.
\item \textit{Overhead} - The percentage of additional messages created upon an incident
detection compared to the total number of messages sent in the network.
\item \textit{Detection Reliability} - Successful detection rate and the false positive
detection rate for various simulation scenarios.
\end{itemize}

We first evaluate the performance of Antilizer with no attack scenario to show that there
is a \textit{zero performance penalty}.
We then evaluate the performance of Antilizer with malicious nodes in the network.
Our results show that our system can detect data loss with high reliability and route
around affected areas.
These changes reduce the loss rate that has been caused by the attacks, as well as the data
delivery delay.

\subsection{Performance Results} \label{subsec:results}

Below, we present the simulation results of our evaluation of Antilizer for varied attack
scenarios and topologies.

\begin{figure}
\centering
\includegraphics[width=\linewidth]{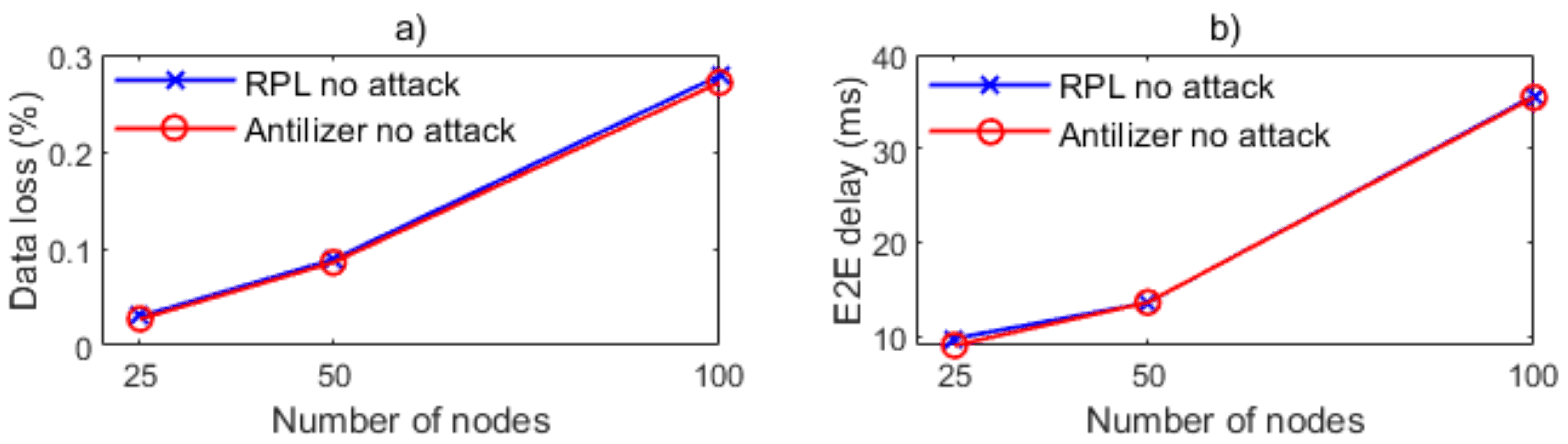}
\caption{Antilizer performance for no attack scenario in $25$, $50$ and $100$ nodes network: (\textbf{a}) Data loss. (\textbf{b}) E2E delay.\vspace{-3mm}}
\label{fig:no_attack}
\end{figure}

\textbf{No Attack Scenario.}
Figure~\ref{fig:no_attack} presents the performance of Antilizer with no attack scenario.
We show that there is a \textit{zero performance penalty} if Antilizer is run in the
network with no attacker.

\begin{figure}
\centering
\includegraphics[width=\linewidth]{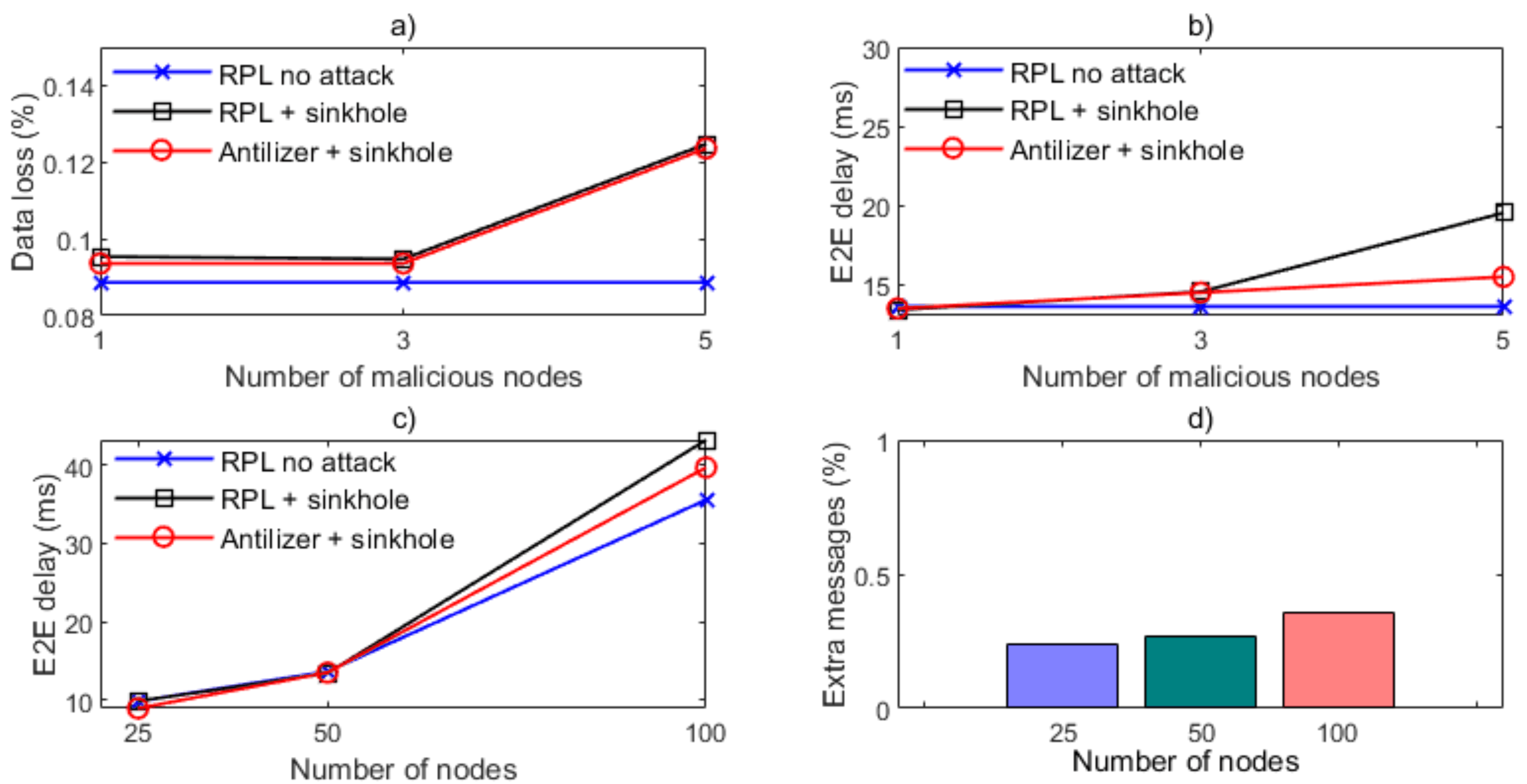}
\caption{Antilizer performance for sinkhole attack: (\textbf{a}) - (\textbf{b}) Data loss and E2E delay ($50$ nodes, multiple attackers) \textbf{c}) - \textbf{d}) Data loss and Overhead ($25$, $50$ and $100$ nodes, single attacker).\vspace{-5mm}}
\label{fig:sinkhole}
\end{figure}

\textbf{Sinkhole Attack.}
Antilizer detects a change in the rank and an increase in receptions and transmissions
during a sinkhole attack.
Once detected, nodes reroute communication whilst triggering the notification scheme.
From the Figure~\ref{fig:sinkhole} we can see that there is no notable impact on E2E data
loss; however routes are compromised so that data is delayed.
Results show that Antilizer keeps the E2E delay in the presence of sinkhole attack close to
normal latency.
The overheads are very low.

\begin{figure}
\centering
\includegraphics[width=\linewidth]{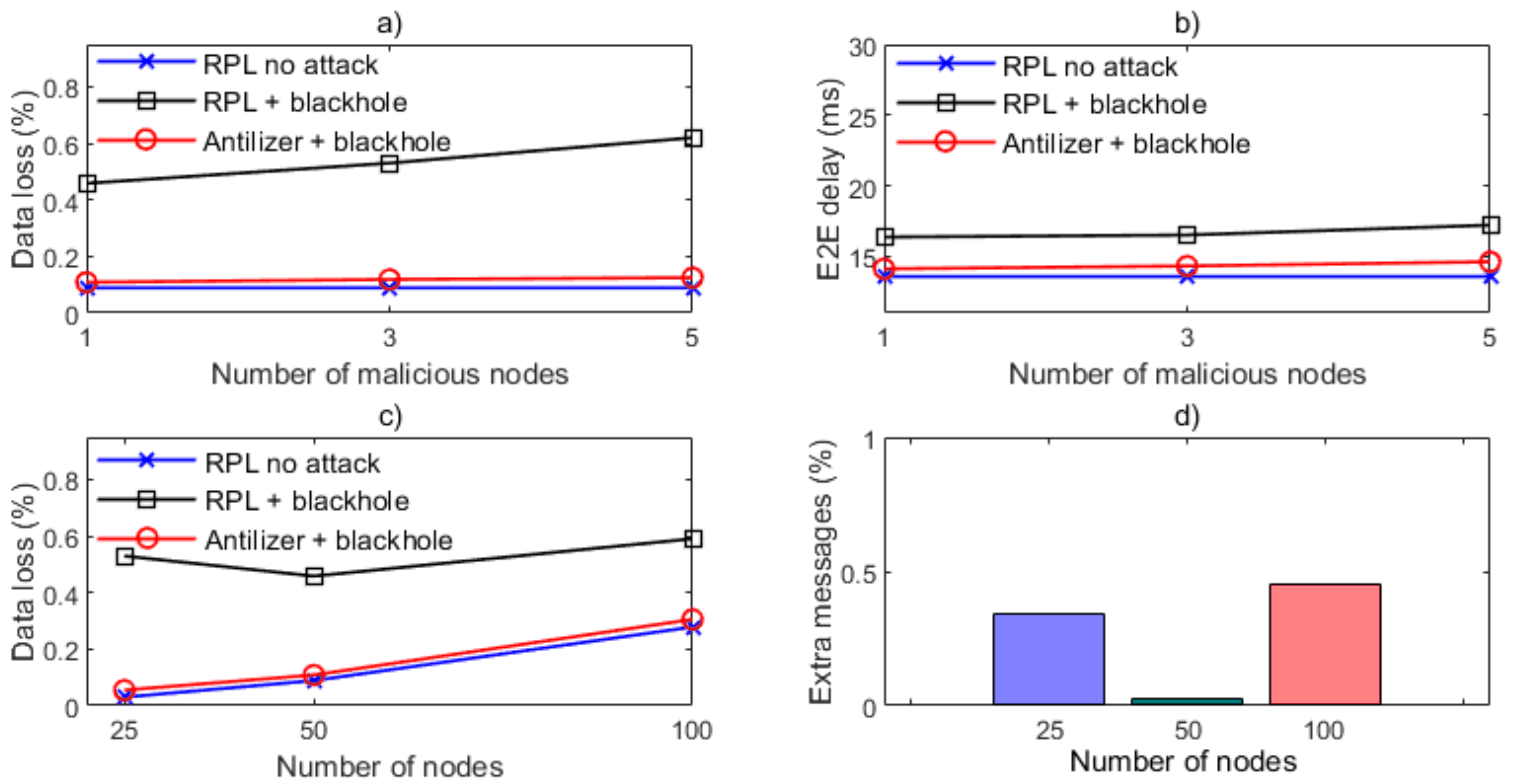}
\caption{Antilizer performance for blackhole attack: (\textbf{a}) - (\textbf{b}) Data loss and E2E delay ($50$ nodes, multiple attackers) \textbf{c}) - \textbf{d}) Data loss and Overhead ($25$, $50$ and $100$ nodes, single attacker).\vspace{-3mm}}
\label{fig:blackhole}
\end{figure}

\textbf{Blackhole Attack.}
During a blackhole attack, Antilizer detects the drop in transmissions and nodes reroute 
data whilst triggering the notification scheme.
Figure~\ref{fig:blackhole} shows that Antilizer reduces the data loss down to 1\%, but
also it keeps the transmission delay closer to normal latency. The overhead introduced is less then 0.5\% across varied attack intensities.

\begin{figure}
\centering
\includegraphics[width=\linewidth]{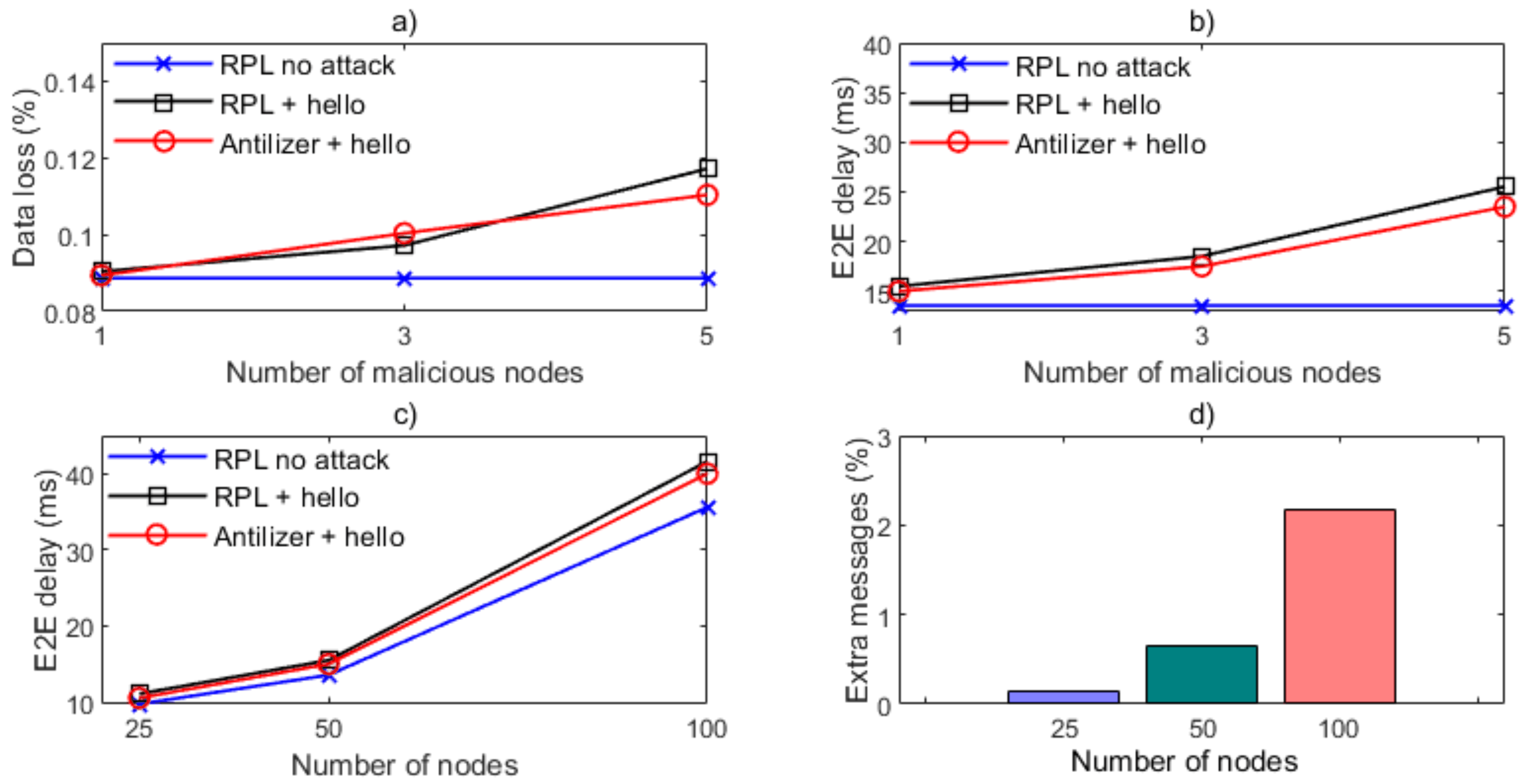}
\caption{Antilizer performance for hello attack: (\textbf{a}) - (\textbf{b}) Data loss and E2E delay ($50$ nodes, multiple attackers) \textbf{c}) - \textbf{d}) Data loss and Overhead ($25$, $50$ and $100$ nodes, single attacker).\vspace{-3mm}}
\label{fig:replay}
\end{figure}

\textbf{Hello Flood Attack.}
Nodes detect an increase in the transmissions of an attacker.
The attack drastically increases the overall number of packets in the network.
The transmission delays increase accordingly as can be seen in Fig.~\ref{fig:replay}.
Due to the nature of the attack, it is not possible to reroute around the affected
regions completely as additional packets are spread across the whole network.
However, all nodes are able to detect the attack and send ANTs to the base station.
The average time needed for a node to detect and report a hello attacker to the base
station is up to two time slots ($40$ seconds).
Therefore, its effects could be prevented by the base station in a very short period.

\begin{figure}
\centering
\includegraphics[width=\linewidth]{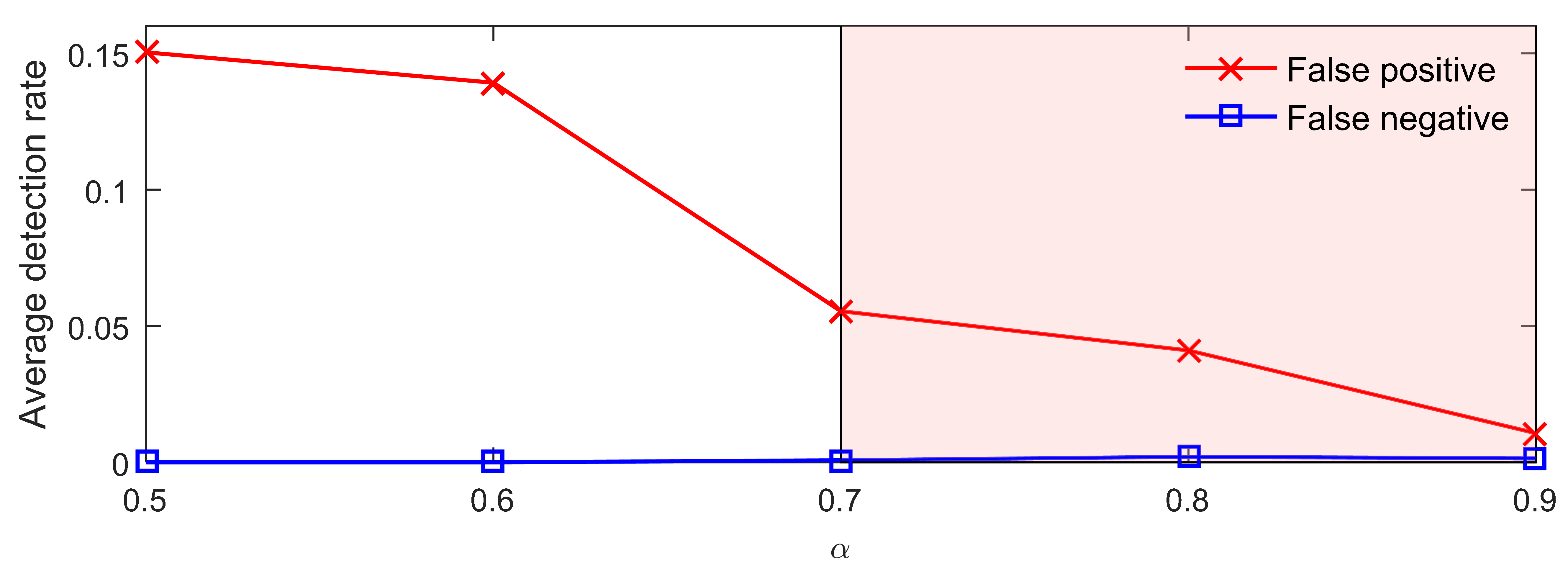}
\caption{Antilizer detection reliability for $\alpha \in [0.5,0.9]$ in $50$ nodes network (average of four attacks, a single attacker).\vspace{-4mm}}
\label{fig:det_reliab}
\end{figure}

\textbf{Detection Reliability.}
The reliability of our detection scheme and its dependence on the choice of
parameter $\alpha$ is shown in Fig.~\ref{fig:det_reliab}.
We performed extensive simulations of three attacks with a single attacking node in a
network of $50$ nodes.
Detection rates were calculated as average rates per node.
The results show that on average for $\alpha \in [0.7,0.9]$ Antilizer can
identify more that 98.6\% anomalies with only around 3.6\% false positive
detections.
While $\alpha=0.9$ gives the lowest number of false positives, we opted for more
conservative approach and $\alpha=0.75$ which ensures a good sensitivity to all
attacks with $99.3$\% detection reliability. 

\section{Conclusions}

In this paper, we presented Antilizer, a novel self-healing scheme to defend
against attacks aimed at network communications. Upon detecting malicious
activities, the system allows data to flow around affected regions so that the
network functionality is not compromised. Our experimental results showed high
effectiveness in terms of data loss rate requiring low operational overheads
for varied attack scenarios.

As part of our future work, we plan to test Antilizer in real test-bed
experiments. Also, we plan to improve the attack diagnosis component by
exploiting reinforcement-learning schemes. This would support the decision
making process at the base station and further improve Antilizer performance.

\bibliographystyle{ACM-Reference-Format}
\bibliography{references}

\end{document}